\PassOptionsToPackage{table}{xcolor}
\documentclass[acmsmall,screen,nonacm]{acmart}

\setcopyright{none}
\usepackage{multirow}
\usepackage{tabularx}
\usepackage{array}
\usepackage{booktabs}
\usepackage{longtable}
\usepackage{makecell}
\usepackage{mdframed}
\usepackage{wrapfig}

\newcommand{\sig}[2]{#1\rlap{$^{#2}$}}
\renewcommand\footnotetextcopyrightpermission[1]{}

\definecolor{verificationrow}{RGB}{232,227,243}

\begin{document}


\title{When CONTRIBUTING Files Fall Short: Revising OSS Contribution Guidelines with Repository Evidence}

\author{Kai Zheng}
\orcid{0009-0008-3707-2477}
\affiliation{%
  \institution{Beihang University}
  \city{Beijing}
  \country{China}
  \postcode{100191}
}
\email{zhengkai@buaa.edu.cn}

\author{XinGuo Zhou}
\orcid{0009-0004-6551-013X}
\affiliation{
  \institution{Beihang University}
  \city{Beijing}
  \country{China}
  \postcode{100191}
}
\email{zhouxinguo@buaa.edu.cn}

\author{Xin Tan}
\orcid{0000-0003-1099-3336}
\affiliation{%
  \institution{Beihang University}
  \city{Beijing}
  \country{China}
  \postcode{100191}
}
\email{xintan@buaa.edu.cn}

\begin{abstract}
Contribution guidelines help newcomers participate in open-source software projects, yet often contain incomplete or outdated information. Existing studies characterize newcomers' information needs and documentation gaps, but revising these guidelines requires recovering and validating project-specific knowledge from scattered sources. We present \textsc{ContribAlign}, a framework for automatically revising \texttt{CONTRIBUTING} files through category-guided evidence integration. A qualitative analysis yields a taxonomy of 30 fine-grained guidance types across six contribution categories. \textsc{ContribAlign} uses a classifier to diagnose coverage of the six categories in the original guidance, then uses the taxonomy to collect and organize project-specific evidence. It filters detected conflicts and uses a large language model to generate category-level revisions. The classifier achieves a weighted F1 score of 0.764, improving on the existing classifier by 0.156, while the verifier detects factual conflicts with 97.3\% precision and 91.5\% estimated recall. Human evaluation shows that \textsc{ContribAlign} achieves the highest overall quality among the original guidelines and three baselines, with significant improvements in usefulness, beginner-friendliness, and completeness. These results suggest that \textsc{ContribAlign} can help maintain more accurate and useful contribution guidelines.
\end{abstract}


\maketitle

\section{Introduction}
In open-source software (OSS) projects, \texttt{CONTRIBUTING} files communicate contribution guidance from maintainers to potential contributors~\cite{classify,gaughan2025introduction}, covering task selection, community interaction, environment setup, code modification and testing, and change submission~\cite{classify,steinmacher2016flosscoach,Elazhary2019Do}. Complete, accurate, and understandable guidance can help newcomers navigate technical and social barriers to their first contributions and prepare well-formed submissions~\cite{steinmacher2015barriers,steinmacher2016flosscoach,githubContributingGuidelines}.

Despite their importance, existing \texttt{CONTRIBUTING} files often provide incomplete guidance and may contain outdated information. Prior research finds that 52\% of the analyzed projects lack guidance addressing at least three of six categories of newcomer onboarding barriers, with common omissions involving task selection and community interaction~\cite{classify,steinmacher2015barriers}. An analysis of the first versions of \texttt{CONTRIBUTING} files further reveals that they primarily describe contribution procedures, while offering limited guidance on community interaction~\cite{gaughan2025introduction}. Moreover, documentation in software projects can become outdated as development practices and project structures evolve. A large-scale analysis of more than 3,000 GitHub projects finds that most projects contain at least one outdated code-element reference during their history~\cite{tan2024detecting}. Consequently, existing \texttt{CONTRIBUTING} files may fail to provide sufficient guidance or accurately reflect the current state of OSS projects~\cite{Elazhary2019Do}.

Keeping these files complete and current requires ongoing attention from maintainers, even when they are familiar with the underlying procedures. With competing development and review responsibilities, maintainers may overlook or postpone necessary updates~\cite{tan2020scaling}. This maintenance effort motivates automated support for improving onboarding documentation. Prior work has explored automatically identifying missing categories in contribution guidelines~\cite{classify} and using large language models (LLMs) to simplify onboarding documentation to reduce information overload~\cite{adejumo2024towards}. However, identifying omissions or simplifying existing text does not provide the information needed to fill gaps or correct inaccurate instructions. These revisions require project-specific evidence from sources such as repository documentation, configurations, and development activities.

Collecting and using this evidence presents two challenges. First, evidence collection must account for the different information needs within contribution guidance~\cite{ju2021onboarding}. Environment setup instructions, for example, may be supported by project guides and configurations, whereas expectations for submitted changes may also be reflected in pull request reviews~\cite{tsay2014let}. Broad retrieval queries may miss specific information needed to revise the guidance, while relying exclusively on documentation may overlook practices reflected in development activities. Second, repository documentation may contain claims that conflict with the current repository state~\cite{devmentorAI}. References to paths, issue labels, or package scripts, for example, may become invalid as a project evolves. Identifying such conflicts before generation can reduce the propagation of outdated instructions, although not all contribution guidance can be verified automatically.

To address these challenges, we propose \textsc{ContribAlign}, a framework for automatically revising existing \texttt{CONTRIBUTING} files through category-guided evidence integration. \textsc{ContribAlign} follows four stages: document diagnosis, evidence collection, evidence verification, and category-wise generation and synthesis. Its key idea is to organize contribution guidance into categories, collect project-specific evidence across six categories, and filter detected factual conflicts before revision. First, \textsc{ContribAlign} uses an improved classifier based on prior work~\cite{classify} to diagnose the coverage of existing guidance by assigning document content to six contribution categories and identifying uncovered categories. Second, it collects taxonomy-guided evidence from project documents and supplements it with direct repository evidence from configurations, collaboration activities, and execution records. Third, it checks selected repository-specific claims and filters fragments containing detected conflicts. Finally, \textsc{ContribAlign} generates guidance for each category and synthesizes the generated content into a coherent document.

To evaluate the effectiveness of the key components and the resulting documents, we conduct both component-level evaluations and human assessment. The classifier achieves a weighted F1 score of 0.764 for guidance coverage diagnosis, while the verifier detects conflicting fragments with 97.3\% precision and 91.5\% estimated recall. In a human evaluation on 20 repositories, the guidelines revised by \textsc{ContribAlign} receive the highest average ratings among the five document versions evaluated for each repository across all four quality dimensions and are selected as the preferred revisions in 33 of 40 comparisons. Overall, this paper makes the following contributions:
\begin{itemize}
    \item We develop a category-aware classifier that automatically diagnoses missing guidance in existing \texttt{CONTRIBUTING} files across six contribution categories.
    \item We design a repository-aware verifier that validates extracted claims against the current repository state to identify conflicting information.
    \item We propose \textsc{ContribAlign}, a framework that automatically revises \texttt{CONTRIBUTING} files by integrating guidance diagnosis, repository evidence, and conflict-aware generation.
\end{itemize}

\section{Background and Related Work}


\subsection{OSS Newcomer Onboarding and Contribution Guidance}

Joining an OSS project requires newcomers to learn both the technical and social aspects of its community~\cite{ducheneaut2005socialization}. Prior studies have identified various barriers that hinder newcomers' first contributions, such as finding suitable tasks, setting up development environments, and communicating with community members~\cite{steinmacher2014preliminary,steinmacher2015barriers}. To alleviate these barriers, researchers have explored different forms of newcomer support, such as recommending suitable tasks~\cite{xiao2022recommending,tan2020first} and providing project-specific onboarding guidance~\cite{steinmacher2016flosscoach}. These studies highlight the importance of accessible and project-specific information in helping newcomers navigate the contribution process.

Among various forms of newcomer support, contribution guidelines provide a repository-level mechanism for communicating project expectations and procedures. Sholler et al.~\cite{sholler2019ten} recommend that OSS projects provide contribution guidelines to help newcomers understand contribution workflows and locate relevant resources. In practice, \texttt{CONTRIBUTING} files serve as the primary carrier of such guidance, covering activities such as task selection, development environment setup, and change submission for review. Therefore, improving the completeness and accuracy of \texttt{CONTRIBUTING} files provides a practical opportunity to enhance newcomer onboarding.

\subsection{Quality and Maintenance of OSS Documentation}

Documentation can be difficult to use because of deficiencies in both content and presentation. Aghajani et al.~\cite{aghajani2019software} analyzed documentation-related artifacts and identified problems involving missing or incorrect information, poor organization, and difficulty finding relevant material. Their subsequent practitioner survey~\cite{aghajani2020software} examined the importance and frequency of documentation issues, including incomplete and outdated information. It also highlighted insufficient time for documentation work and the need for better automation.

For contribution documentation, limitations can arise when the files are first introduced. Gaughan et al.~\cite{gaughan2025introduction} examined initial README and \texttt{CONTRIBUTING} files and found an emphasis on technical information and procedures, with limited attention to community development. As projects evolve, existing instructions may also need revision to remain applicable. Supporting contribution guidelines therefore involves assessing both whether they cover newcomers' information needs and whether their content remains consistent with the project.

\subsection{Automated Documentation Analysis and Generation}

Automated approaches have been explored to identify gaps and inconsistencies in software documentation. Fronchetti et al.~\cite{classify} trained a classifier to categorize paragraphs in \texttt{CONTRIBUTING} files according to six newcomer onboarding categories and found that 52\% of the examined files lacked guidance in at least three categories. However, this approach only diagnoses coverage gaps without assessing whether existing instructions are correct or actionable. Similarly, Tan et al.~\cite{tan2024detecting} detected outdated code-element references in documentation, helping identify potential maintenance needs but not generating revised guidance.

Recent LLM-based approaches have been applied to improve software documentation. LARCH~\cite{larch} selects representative source files using heuristics and weak supervision for README generation, demonstrating the importance of selecting appropriate project context for generating useful documentation. However, contribution guidance requires information beyond source code, including project procedures, configurations, and collaboration practices. Other studies focus on improving existing onboarding documents: Adejumo and Johnson~\cite{adejumo2024towards} used LLMs to simplify \texttt{CONTRIBUTING} files and improve readability, while Feng et al.~\cite{feng2026restructure} introduced VisDoc to restructure onboarding documents into interactive task representations. Nevertheless, these approaches primarily improve document expression and organization, without addressing whether the underlying guidance is complete, up-to-date, and consistent with repository facts.

Taken together, existing studies provide important support for documentation improvement but leave the challenge of reliably revising contribution guidelines. \textsc{ContribAlign} addresses this challenge by diagnosing missing guidance, collecting and verifying repository-specific evidence, and filtering detected conflicts before generating revised \texttt{CONTRIBUTING} files.

\section{Design Foundation}\label{sec:design_foundation}
This section establishes the design foundation for \textsc{ContribAlign} by characterizing the information needs that guide evidence collection and verification. We first introduce the existing classification framework for contribution guidance, then describe our qualitative analysis, and finally present the resulting fine-grained taxonomy.

\subsection{Existing Classification Framework}

Prior work organizes the guidance in \texttt{CONTRIBUTING} files into six contribution categories: \textit{Talk to the Community}, \textit{Choose a Task}, \textit{Contribution Flow}, \textit{Build Local Workspace}, \textit{Deal with the Code}, and \textit{Submit the Changes}~\cite{classify}. These categories were operationalized by Fronchetti et al. based on Steinmacher et al.'s model of newcomer onboarding barriers~\cite{steinmacher2015barriers,steinmacher2016flosscoach}.
The six categories describe broad contribution tasks but do not distinguish the specific information provided within each task. For example, \textit{Build Local Workspace} may include tool versions, dependency installation, and build instructions. These fine-grained types may require different repository sources and verification procedures. To characterize this finer-grained structure, we qualitatively analyzed the guidance within each of the six categories.

\subsection{Qualitative Analysis}

We adopted open coding~\cite{corbin1990grounded}, following Prana et al.'s approach to categorizing GitHub documentation~\cite{prana2019categorizing}, to derive fine-grained guidance types within the six contribution categories.
We used the manually annotated dataset of Fronchetti et al.~\cite{classify}, which contains English \texttt{CONTRIBUTING} files from 500 high-starred GitHub repositories across ten programming languages. After excluding paragraphs labeled as \textit{No Category Identified}, the dataset contains 13,260 paragraph records across the six categories. We analyzed each category separately, allowing each paragraph to map to multiple fine-grained types or none. The first two authors independently performed open coding on 50 \texttt{CONTRIBUTING} files, with no new types emerging toward the end of the coding process, indicating data saturation.

The two authors then consolidated the initial codes, merged semantically overlapping codes, and refined their definitions and boundaries to establish the codebook. To assess its reliability, they independently annotated another 50 \texttt{CONTRIBUTING} files, achieving a Cohen's \(\kappa\) of 0.74. Disagreements were adjudicated through discussion with a third author experienced in qualitative analysis. The resulting fine-grained types form the basis for category-aware evidence retrieval and processing in \textsc{ContribAlign}.

\subsection{Fine-Grained Taxonomy}
The analysis yielded 30 fine-grained guidance types across the six categories, summarized in Table~\ref{tab:fine-grained-taxonomy}. \textit{Talk to the Community} concerns participation in community discussions and obtaining support. \textit{Choose a Task} covers discovering and selecting tasks, coordinating before starting work, and understanding the scope of acceptable contributions. Unlike the other categories, which are organized primarily by guidance content, the types within \textit{Contribution Flow} follow the sequence of contribution activities, from initiating a contribution to submitting a change. \textit{Build Local Workspace} concerns preparing and validating the development environment. \textit{Deal with the Code} covers the conventions, quality requirements, and technical constraints for code changes. \textit{Submit the Changes} concerns the organization and submission of changes and the subsequent review and merging. This taxonomy specifies the information to seek within each category and guides the evidence collection of \textsc{ContribAlign} (Section~\ref{sec:approach}).

\begin{table}[t]
\centering
\caption{Fine-Grained Taxonomy of Contribution Guidance}
\label{tab:fine-grained-taxonomy}
\footnotesize
\setlength{\tabcolsep}{5pt}
\renewcommand{\arraystretch}{1.08}

\begin{tabularx}{\columnwidth}{
    >{\raggedright\arraybackslash}p{0.25\columnwidth}
    >{\raggedright\arraybackslash}X
}
\toprule
\textbf{Contribution Category} &
\textbf{Fine-Grained Types} \\
\midrule

Talk to the Community &
Access to Community Channels; Channel Use; Help-Seeking; Community Interaction Norms; Code of Conduct \\
\addlinespace[2pt]

Choose a Task &
Task Discovery and Selection; Contribution Needs and Priorities; Task Coordination; Contribution Scope and Acceptance Boundaries \\
\addlinespace[2pt]

Contribution Flow &
Pre-implementation Coordination; Repository and Branch Preparation; Change Implementation; Pre-submission Validation; Submission and Review \\
\addlinespace[2pt]

Build Local Workspace &
Toolchain Requirements; Dependency and Service Setup; Project Build and Installation; Local Development Execution; Workspace Readiness Checks; Platform Configuration and Alternative Environments \\
\addlinespace[2pt]

Deal with the Code &
Coding Conventions; Code Design; Code Documentation; Code Quality Checks; Technical Constraints \\
\addlinespace[2pt]

Submit the Changes &
Pull Request Content; Commit Organization; Submission Validation and Artifacts; Contribution Licensing and Sign-off; Review and Merge \\

\bottomrule
\end{tabularx}
\end{table}

\section{ContribAlign}\label{sec:approach}

\begin{figure*}[t]
    \centering
    \includegraphics[width=\textwidth]{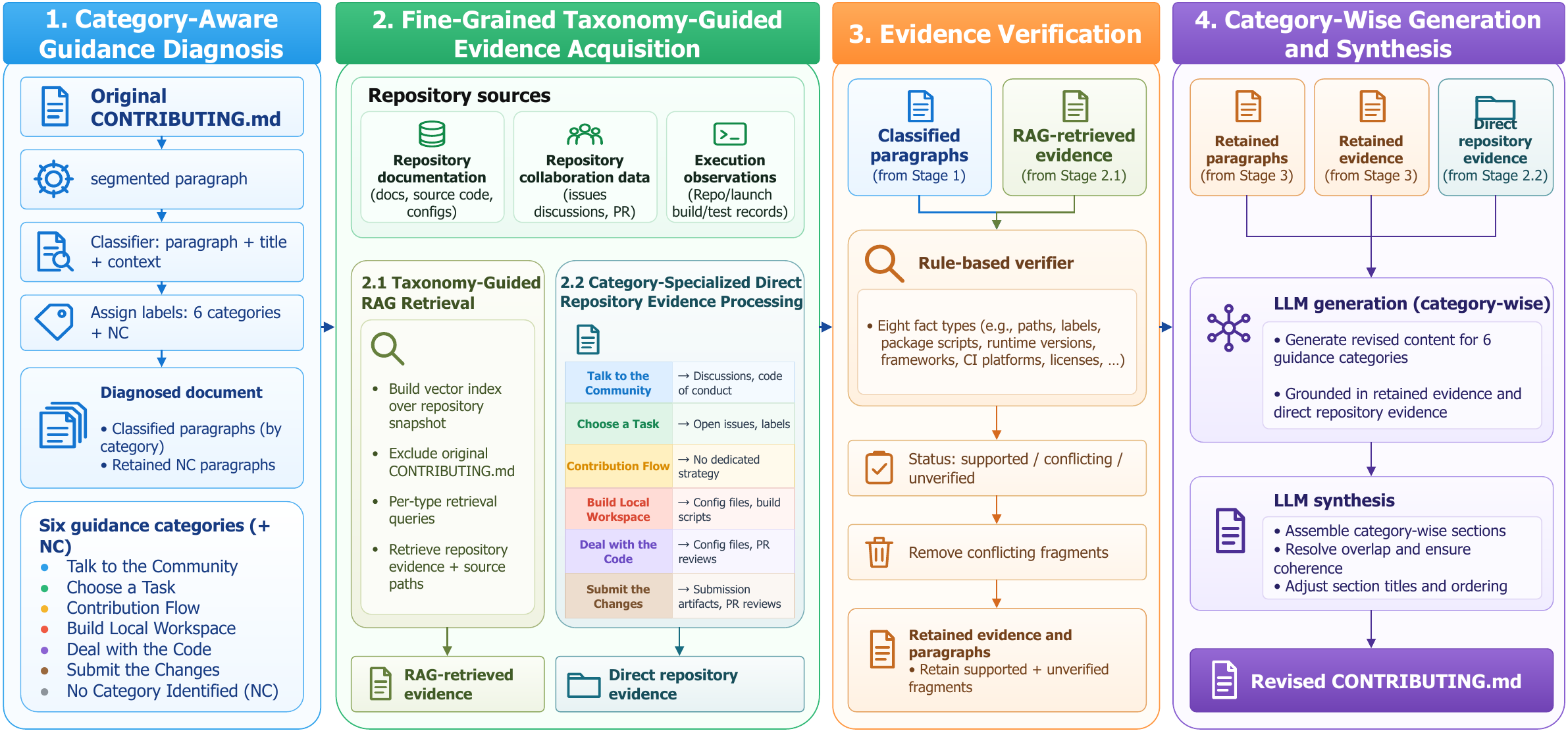}
    \caption{Overview of ContribAlign}
    \label{fig:method-overview}
    \vspace{-0.2cm}
\end{figure*}

As shown in Figure~\ref{fig:method-overview}, \textsc{ContribAlign} revises existing \texttt{CONTRIBUTING} files through four stages. First, it classifies the original content into six contribution categories defined in prior work~\cite{classify} and identifies potential coverage gaps. It then collects repository evidence to supplement and revise the guidance, using the fine-grained taxonomy developed in Section~\ref{sec:design_foundation} to specify the information needed within each category. Next, it checks selected repository-specific claims in the original guidance and retrieved evidence and removes fragments containing detected conflicts. Finally, it uses the retained content and direct repository evidence to generate category-specific drafts and synthesize them into a coherent document. The following subsections describe the four stages.

\subsection{Category-Aware Guidance Diagnosis}
\label{sec:diagnosis}
\textsc{ContribAlign} first diagnoses the coverage of the existing \texttt{CONTRIBUTING} file using the six contribution categories proposed by Fronchetti et al.~\cite{classify}. These categories provide a structured basis for identifying missing guidance and organizing existing content for category-wise revision. Since their original classifier relies on traditional machine learning and classification directly affects subsequent diagnosis, we develop a pre-trained language model-based classifier, with its model and input configuration determined through the evaluation in Section~\ref{sec:experiment_setup_RQ1}.

We parse each \texttt{CONTRIBUTING} file using a CommonMark parser with GitHub Flavored Markdown extensions and segment it into paragraphs while preserving headings and neighboring paragraphs as contextual information. The classifier assigns each paragraph to one of the six categories or \textit{No Category Identified}.
We then group paragraphs by their predicted categories to construct a coverage profile. A category is considered covered if at least one paragraph is assigned to it and missing otherwise; coverage indicates only the presence of relevant content, not its adequacy or completeness. The grouped paragraphs serve as the original content for category-wise revision, while missing categories are supplemented with collected evidence. Paragraphs classified as \textit{No Category Identified} are retained for supplementary processing.

\subsection{Fine-Grained Taxonomy-Guided Evidence Acquisition}
In the second stage, \textsc{ContribAlign} collects evidence for all six contribution categories, guided by the fine-grained taxonomy in Section~\ref{sec:design_foundation}. Because information needs vary across the contribution process~\cite{steinmacher2015barriers,steinmacher2015Social}, we combine taxonomy-guided retrieval for retrieval-augmented generation (RAG) with category-specialized processing of direct repository evidence. The RAG component retrieves relevant content from a shared index of the repository snapshot using queries derived from the fine-grained types, while direct processing collects category-specific evidence from repository configurations, collaboration activities, and execution observations. Together, these sources provide project-specific evidence for subsequent category-wise revision.

\subsubsection{Taxonomy-Guided RAG Retrieval}
\label{sec:rag-retrieval}
RAG-based retrieval collects evidence from the repository snapshot for each contribution category. To ensure that every category retrieves the information it needs, \textsc{ContribAlign} derives the retrieval queries from the fine-grained taxonomy: each fine-grained type that requires document retrieval is associated with a dedicated query, and the retrieval terms are defined by the taxonomy itself. For example, the \textit{Choose a Task} category is served by four dedicated queries, one for each of its fine-grained types. For instance, the \textit{Task Discovery and Selection} type is associated with a keyword-based retrieval query focusing on how contributors find tasks, with terms such as ``good first issue'' and ``issue tracker''.

To support these queries, \textsc{ContribAlign} builds a vector index over the entire repository snapshot. Every readable text file in the snapshot is converted into a text record, and we retain the file path as metadata to attribute the source when organizing evidence. All records are chunked, embedded, and stored in a vector database. The original \texttt{CONTRIBUTING} file is excluded from this index. It is processed separately in the first stage and enters the downstream pipeline as the original content to be revised.
\textsc{ContribAlign} executes each query against the shared index and associates every retrieved text block with the corresponding fine-grained type and source file path. If the same text block is retrieved for multiple fine-grained types within a category, \textsc{ContribAlign} retains a single copy and records all associated types. Evidence for different categories is organized independently, so a text block relevant to multiple categories is retained in each corresponding category.

\subsubsection{Category-Specialized Direct Repository Evidence Processing}
\label{sec:direct-evidence}
In addition to taxonomy-guided RAG retrieval, \textsc{ContribAlign} applies category-specialized processing to collect direct repository evidence from configurations, collaboration data, and execution records. The evidence is organized by contribution category with its source information retained. Five categories receive dedicated processing. \textit{Contribution Flow} uses retained original guidance, when available, and evidence retrieved for that category to capture the overall contribution sequence.

\textbf{Talk to the Community.}
\textsc{ContribAlign} supplements community guidance with GitHub Discussions and the repository's Code of Conduct. When Discussions is enabled, it collects the access URL and category descriptions to guide contributors toward appropriate channels for questions and discussions. It also retrieves the local Code of Conduct, retaining its path and full content to support guidance on expected behavior and reporting procedures~\cite{li2021code}.


\textbf{Choose a Task.}
\textsc{ContribAlign} combines the repository's open issues and complete label catalog, retaining only labels used on open issues. For each retained label, it records the name, official description, and associated open-issue count, and ranks labels by this count. This evidence supports task-selection guidance grounded in the repository's current open tasks.

\textbf{Build Local Workspace.}
This category helps newcomers set up a local development environment and build, run, and test the project. \textsc{ContribAlign} collects relevant configuration files, retaining their paths and complete contents as evidence of tool requirements and setup procedures. Since these files may omit necessary steps or command ordering, we supplement them with execution records from RepoLaunch~\cite{repolaunch}, which supports automated building and testing across multiple languages and platforms.
RepoLaunch installs dependencies and builds the repository in a container, checks that most tests and core functionality tests pass, and re-executes the extracted rebuild commands. This combines project-declared requirements with commands validated in the tested environment, providing a stronger basis for actionable setup guidance. \textsc{ContribAlign} retains rebuild and test commands from successful runs and includes tool versions only when explicitly reported. The workflow is not executed for unsupported languages, and only successful build and test records are retained as evidence. 

\textbf{Deal with the Code.}
To help newcomers follow project-specific expectations when modifying and submitting code, \textsc{ContribAlign} supplements document retrieval with two complementary evidence sources: project configuration files and pull request review records. Configuration files specify development tools and settings, while review records capture expectations expressed during the review process.
For configuration evidence, the system collects files related to \textit{Coding Conventions}, \textit{Code Quality Checks}, and \textit{Code Documentation}, covering formatting rules, quality-check settings, and documentation tools. Their repository paths and contents are retained as direct evidence for guidance generation.
For review evidence, the system excludes content from pull request authors and bot accounts, then provides the remaining records and the fine-grained type definitions under \textit{Deal with the Code} and \textit{Submit the Changes} to the LLM in a single call. The LLM extracts recurring review expectations, filters out suggestions specific to individual changes, and assigns each expectation to a fine-grained type. These expectations serve as supplementary evidence for guidance generation.

\textbf{Submit the Changes.}
When submitting changes, newcomers need to understand the project-specific requirements for submission and review. These requirements are reflected not only in repository documents but also in submission templates, related configurations, and recurring review practices. To this end, \textsc{ContribAlign} collects three groups of repository artifacts: pull request templates, commit message templates and \texttt{commitlint} configurations, and documents related to the Developer Certificate of Origin (DCO) and Contributor License Agreement (CLA). \textsc{ContribAlign} retains the repository paths and contents of these files as direct evidence for submission guidance.

\textsc{ContribAlign} also retains the review requirements assigned to \textit{Submit the Changes} in the LLM call described above, together with their fine-grained types. These requirements are especially helpful for \textit{Submission Validation and Artifacts} and \textit{Review and Merge}, where repository documents often provide little explicit guidance. Together with the repository artifacts described above, these review requirements serve as evidence for generating submission guidance.

\subsection{Evidence Verification}
\label{sec:verify}
As repositories evolve, repository documentation may become inconsistent with the current repository state and retain outdated references to repository entities~\cite{aghajani2019software,tan2024detecting}. To reduce the propagation of such stale information into the generated contribution guidelines, we design a rule-based evidence verifier that checks automatically verifiable repository-specific claims against current repository state. The verifier processes two inputs: paragraphs retained from the original \texttt{CONTRIBUTING} file during document diagnosis (Section~\ref{sec:diagnosis}) and text chunks retrieved by RAG (Section~\ref{sec:rag-retrieval}). This verification step focuses on potentially stale claims in the original guidance and RAG-retrieved evidence. Direct repository evidence follows a separate, category-specific processing path into generation.

\textsc{ContribAlign} first segments the original guidance and RAG-retrieved evidence into fragments using paragraph, list-item, and code-block boundaries where applicable. A fragment containing multiple independently verifiable statements is further divided, allowing a conflicting statement to be removed without discarding the remaining content. \textsc{ContribAlign} then identifies repository-specific claims that can be automatically checked, such as references to repository paths, issue labels, package scripts, and runtime versions. We define type-specific patterns and parsers to extract the repository entities or values referenced in these claims, then compare them with the current repository snapshot, repository metadata, configuration files, and other structured information collected during evidence acquisition. The verifier covers eight types of such claims, as summarized in Table~\ref{tab:verification-rules}. Verification is limited to claims whose validity can be determined from the available repository evidence. Each fragment is assigned one of three statuses:
\begin{itemize}
    \item \textit{supported}: all identified claims can be confirmed against the current repository state;
    \item \textit{conflicting}: at least one identified claim clearly contradicts the current repository state;
    \item \textit{unverified}: the fragment contains no automatically verifiable claims, or the available repository information is insufficient to determine their validity.
\end{itemize}

Only \textit{conflicting} fragments are removed. \textit{Supported} and \textit{unverified} fragments are retained for category-wise generation. A \textit{supported} label indicates that the claims identified and checked by the verifier are supported by repository evidence; it does not imply that every statement in the fragment has been verified. The verifier therefore filters detected factual conflicts within its supported claim types rather than establishing the correctness of all retained content.

\begin{table*}[t]
\centering
\caption{Repository-Specific Claims and Verification Rules}
\label{tab:verification-rules}
\footnotesize
\setlength{\tabcolsep}{5pt}
\renewcommand{\arraystretch}{1.08}

\begin{tabularx}{\textwidth}{
    >{\raggedright\arraybackslash}p{0.18\textwidth}
    >{\raggedright\arraybackslash}X
}
\toprule
\textbf{Claim Type} & \textbf{Verification Rule} \\
\midrule

Repository paths &
Resolve relative file and directory references and check them against
the repository inventory. Missing references or case mismatches indicate
conflicts; paths potentially generated during building or installation
remain unverified. \\
\addlinespace[2pt]

Issue labels &
Match explicitly named labels against the repository's label catalog,
ignoring case. Missing labels indicate conflicts. \\
\addlinespace[2pt]

Package scripts &
Check invoked npm, Yarn, or pnpm scripts against the \texttt{scripts}
field of the applicable \texttt{package.json}. Undefined scripts indicate
conflicts; unresolved package scopes remain unverified. \\
\addlinespace[2pt]

Runtime versions &
Compare stated runtime versions with constraints in project manifests
and toolchain configurations. Incompatible versions indicate conflicts;
unresolved project scopes or version relationships remain unverified. \\
\addlinespace[2pt]

Target frameworks &
Check stated .NET frameworks against \texttt{TargetFramework} or
\texttt{TargetFrameworks} declarations in project files.
Undeclared frameworks indicate conflicts. \\
\addlinespace[2pt]

CI platforms &
Compare named CI platforms with repository configurations.
Explicit contradictory evidence indicates a conflict;
missing configurations alone leave the claim unverified. \\
\addlinespace[2pt]

Governance files &
Check references to pull request templates and Codes of Conduct against
the repository inventory and recognized standard locations.
Missing referenced files indicate conflicts. \\
\addlinespace[2pt]

Licenses &
Compare stated repository or contribution licenses with the repository's
license files. Contradictions indicate conflicts;
ambiguous or unidentified licenses remain unverified. \\

\bottomrule
\end{tabularx}
\end{table*}

\subsection{Category-Wise Generation and Synthesis}
In the final stage, \textsc{ContribAlign} generates guidance separately for the six contribution categories. The input for each category includes relevant paragraphs from the original \texttt{CONTRIBUTING} file retained after verification, if any, and two types of category-specific evidence: retained evidence retrieved through RAG and direct repository evidence. \textsc{ContribAlign} organizes this evidence according to the fine-grained taxonomy but does not require the generated content to contain a separate subsection for every fine-grained type.

For each category with available evidence, \textsc{ContribAlign} makes an independent LLM call. If the original \texttt{CONTRIBUTING} file already contains relevant content for the category, the prompt asks the LLM to revise and supplement the existing guidance using the collected evidence. If no original content is available, the prompt asks the LLM to generate new guidance from the evidence. When the supplied evidence is insufficient to support category-specific guidance, the LLM may return an omission signal instead of generating a draft. Only generated category drafts are passed to the final synthesis stage.

To avoid discarding potentially useful original guidance that is not assigned to any of the six categories, \textsc{ContribAlign} separately processes paragraphs classified as \textit{No Category Identified} and retained after verification. An additional LLM call determines whether these paragraphs contain contribution guidance not already covered by the category drafts. Such content is organized into a supplementary draft when applicable; otherwise, it is omitted.

Finally, \textsc{ContribAlign} uses an additional LLM call to synthesize the generated category drafts and the optional supplementary draft into a complete \texttt{CONTRIBUTING} file. The synthesis prompt allows the model to adjust the number, titles, and order of sections and to merge duplicated content across drafts, while instructing it not to introduce new repository-specific information beyond the supplied drafts. 

\section{Evaluation}
\subsection{Research Questions}
\textsc{ContribAlign} relies on two key capabilities before generating contribution guidelines: diagnosing which contribution categories are already covered by the existing \texttt{CONTRIBUTING} file and verifying repository-specific claims in the original guidance and RAG-retrieved evidence. Errors in either step may affect the evidence used for generation. We therefore evaluate these two intermediate capabilities separately and then assess the quality of the final generated documents. Specifically, we investigate the following research questions:

\textbf{RQ1: How accurately can \textsc{ContribAlign} classify existing contribution guidance into contribution categories?}
Since paragraph classification underlies category-aware guidance diagnosis, we evaluate multiple pre-trained language models to identify a reliable classifier for \textsc{ContribAlign}. We compare them with the traditional machine-learning classifier proposed by Fronchetti et al.~\cite{classify} and examine the effect of incorporating headings and surrounding context.

\textbf{RQ2: How accurately can \textsc{ContribAlign} verify repository-specific claims in the original guidance and RAG-retrieved evidence?}
We compare the rule-based verifier with manual annotations and measure its accuracy in classifying fragments as \textit{supported}, \textit{conflicting}, or \textit{unverified}.

\textbf{RQ3: How does the quality of contribution guidelines revised by \textsc{ContribAlign} compare with existing and baseline-generated guidelines?}
We conduct a human evaluation to compare the candidate documents in terms of completeness, correctness, beginner-friendliness, and usefulness.

\subsection{Experimental Setup}
\subsubsection{RQ1: Contribution Guidance Classification}


\label{sec:experiment_setup_RQ1}

\paragraph{Dataset.}
We use the manually annotated dataset released by Fronchetti et al.~\cite{classify}, which contains contribution-guidance content collected from 500 OSS repositories and annotated with six contribution categories. Following the preprocessing described in Section~\ref{sec:diagnosis}, we obtain 20,704 paragraph-level instances.
We split the dataset at the repository level to prevent paragraphs from the same repository from appearing in different partitions. Repositories are assigned to the training, validation, and test sets using a greedy iterative procedure that aims to preserve both the target partition sizes and the overall category distribution. The resulting training, validation, and test sets contain 14,786, 1,765, and 4,153 paragraphs, respectively, corresponding to 71.42\%, 8.52\%, and 20.06\% of the dataset.

\paragraph{Compared Models.}
We compare five pre-trained language models: BERT-base~\cite{devlin2019bert}, ALBERT~\cite{lan2019albert}, RoBERTa~\cite{liu2019roberta}, DeBERTa-v3-large~\cite{he2021debertav3}, and CodeBERT~\cite{feng2020codebert}. The first four represent widely used general-purpose language models, while CodeBERT is pre-trained on both natural-language and programming-language data and is therefore relevant to software-engineering text.
We use the LinearSVC classifier proposed by Fronchetti et al.~\cite{classify} as the baseline. Because our repository-level split differs from that used in their original study, we retrain and evaluate LinearSVC on our data partitions using their reported configuration. This ensures that all models are evaluated on the same held-out test set.

\paragraph{Input Configurations.}
To examine whether document context improves classification, we consider three input configurations: the paragraph alone (\texttt{paragraph}); the paragraph prefixed with its heading (\texttt{paragraph+title}); and the paragraph supplemented with its heading and neighboring paragraphs under the same heading (\texttt{paragraph+title+context}). For the third configuration, preceding and following paragraphs are added alternately, starting with the nearest neighbors, subject to a maximum input length of 256 tokens. All configurations use the same paragraph labels and data partitions.

\paragraph{Evaluation and Model Selection.}
We use weighted F1 as the primary evaluation metric because it accounts for differences in category frequency when summarizing multiclass classification performance. Fronchetti et al.~\cite{classify} also report weighted F1, which facilitates comparison with their classifier. We additionally report precision and recall.
For the pre-trained models, model and input-configuration selection is performed exclusively on the validation set. The model--configuration combination achieving the highest weighted F1 is selected as the paragraph classifier used in \textsc{ContribAlign}. Its performance is then reported on the held-out test set.

\subsubsection{RQ2: Evidence Verification}

We evaluate the rule-based verifier using fragments from the original \texttt{CONTRIBUTING} files and RAG-retrieved evidence across the 94 repositories described in Section~\ref{sec:experiment_setup_RQ3}. After merging duplicate fragments within each repository, the verifier identified 1,537 checkable candidate fragments, classifying 1,239 as \textit{supported}, 37 as \textit{conflicting}, and 261 as \textit{unverified}.
We stratified the fragments by verifier-assigned status and determined the sample sizes at a 95\% confidence level with a $\pm$5\% margin of error. We sampled 294 of the 1,239 \textit{supported} fragments and 156 of the 261 \textit{unverified} fragments, and included all 37 \textit{conflicting} fragments, yielding 487 distinct fragments from 74 repositories.

The first author manually annotated the sampled fragments following predefined criteria. The annotation focuses only on the elements identified and checked by the verifier. For each fragment, the first author interprets these elements in context, cross-checks them against the corresponding repository snapshot, assigns a \textit{supported}, \textit{conflicting}, or \textit{unverified} label, and records the evidence supporting the judgment. This evidence-based procedure makes each annotation traceable to the repository snapshot. Using the human labels as ground truth, we construct a three-class confusion matrix. To account for stratified sampling, we weight each row by the ratio of its stratum population to sample size before computing overall accuracy and per-class precision, recall, and F1. We further analyze false \textit{conflicting} predictions and actual conflicts missed by the verifier.


\subsubsection{RQ3: Generated Guideline Quality}
\label{sec:experiment_setup_RQ3}
\paragraph{Dataset}
For RQ3, we select repositories from the RQ1 test set to keep the revision evaluation independent of classifier training and model selection. This set contains 111 repositories from the annotated \texttt{CONTRIBUTING} corpus of Fronchetti et al.~\cite{classify}. We exclude one inaccessible repository, two without retrievable current \texttt{CONTRIBUTING} files, and 14 archived repositories, leaving 94 repositories. Because projects may have evolved since the original corpus was collected, we retrieve the current \texttt{CONTRIBUTING} files. We also collect repository information and download snapshots of the default branches through GitHub APIs to obtain up-to-date evidence for revision.

We collect repository data to provide project context and evidence for revising contribution guidelines. Repository metadata, including descriptions, primary languages, default branches, and topics, establish the project context. Repository labels and all issues open at collection time, including their numbers and labels, support guidance on finding suitable tasks. To capture contribution practices and review expectations, we collect the 50 most recently created pull requests per repository, regardless of status (open, closed, or merged), together with their comments, reviews, and review comments. We also record whether Discussions are enabled and, where available, their categories to identify community communication channels.

\paragraph{Baselines}
In addition to comparing the generated documents with the original, human-written \texttt{CONTRIBUTING} files, we implement three generation baselines. All three use the same revision prompt, which asks the LLM to revise the original document and cover the six predefined contribution categories. They differ in the additional repository context provided to the LLM. The full prompt is available in our replication package.

\begin{itemize}
    \item \textbf{Direct LLM.} This baseline revises the existing \texttt{CONTRIBUTING} file without additional repository evidence retrieval. Its input consists of the original \texttt{CONTRIBUTING} file, repository metadata (repository name, description, and primary language), and the README file.

    \item \textbf{Adapted LARCH.} LARCH~\cite{larch} is an LLM-based system that generates README files from Python repositories by selecting representative code and prompting an LLM with it. We adapt it to generate \texttt{CONTRIBUTING} files and extend its code selection procedure to support other programming languages. The selected repository context, together with the original contribution guidance, serves as the additional input to the revision prompt.

    \item \textbf{Generic RAG.} This baseline uses the same retrieval corpus and index as \textsc{ContribAlign} and retrieves evidence with a single query that lists the names of the six contribution categories. The query returns the top 10 candidate fragments, a budget following common settings in existing studies~\cite{chen2026sweexpexperiencedrivensoftwareissue,wu2026doeschunking} and accounting for the effect of retrieval count on the length of the generated document. 
\end{itemize}

\paragraph{Base LLM}
We use DeepSeek-V4.1-Flash~\cite{xu2026deepseek} as the underlying LLM for \textsc{ContribAlign} and all baseline methods, thereby controlling for differences in model capability across approaches. DeepSeek-V4.1-Flash supports a 1M-token context window~\cite{deepseek2026pricing}, and DeepSeek reports that it outperforms V4-Pro on multiple benchmarks while offering lower inference costs~\cite{deepseek2026v41flash,deepseek2026pricing}. We therefore select it to balance model capability and experimental cost. All generation calls use a temperature of 0.2 and a top-p of 0.8.

\paragraph{Human Evaluation Protocol.}
RQ3 compares the original \texttt{CONTRIBUTING} files with those revised by \textsc{ContribAlign} and three baselines. Because automated metrics cannot adequately capture their factual accuracy and practical usefulness for newcomers, we conducted a human evaluation on 20 repositories, stratified by primary programming language from the 94 targets. The study was approved by our university's institutional ethics review committee. Before the formal evaluation, two participants piloted the procedure to clarify the criteria and scoring process; their results were excluded from the analysis.

We recruited 10 graduate students with software engineering backgrounds who had not previously contributed to the evaluated projects to participate in the formal evaluation. Their technical backgrounds enabled them to examine repository evidence, while their lack of prior contribution experience in these projects provided a perspective relevant to project newcomers. We randomly assigned four repositories to each evaluator. Each repository was independently evaluated by two evaluators, and their ratings were averaged for analysis. For each repository, the evaluator received a shared scoring guide and a downloaded repository snapshot with the original \texttt{CONTRIBUTING} file removed. The evaluator reviewed five candidate documents: the original human-written file and the versions generated by \textsc{ContribAlign} and the three baselines. Each evaluator thus assessed 20 documents against repository evidence across four quality dimensions. Having the same evaluator assess all five documents for a repository supported comparisons under a consistent evaluation standard. The documents were anonymized and presented in random order to reduce potential source and presentation-order bias.

Drawing on prior work on software documentation quality and the information needs of OSS newcomers~\cite{alchimowicz2016coca,classify,steinmacher2015barriers}, we evaluated each candidate document along four dimensions:
\begin{itemize}
\item \textbf{Correctness:} the extent to which the document accurately describes commands, paths, versions, project policies, and contribution procedures in light of current repository evidence.
\item \textbf{Usefulness:} the extent to which the document helps contributors find tasks, communicate with the community, set up the development environment, modify and validate code, and submit contributions.
\item \textbf{Beginner-friendliness:} how easily contributors unfamiliar with the project can understand the document, find the information they need, and follow its instructions.
\item \textbf{Completeness:} the extent to which the document covers the information needed to contribute. Only accurate and practically useful content counts toward coverage.
\end{itemize}
Evaluators closely read each document and cross-checked factual statements against relevant repository files, including whether referenced paths existed, commands matched available scripts, and requirements agreed with configuration files and project documentation. They also examined whether instructions explained prerequisites, provided actionable steps in a coherent order, and omitted information needed to complete contribution tasks. This process combined statement-level verification with contextual judgments about the document as a whole.

Each document received a five-point Likert rating on each dimension~\cite{likert1932technique}, from 1 (the criterion is not met at all) to 5 (the criterion is fully met), yielding 200 document assessments and 800 dimension-level ratings. 
After scoring the five documents for a repository, each evaluator selected the document they would most prefer to provide to newcomers. An \textit{Optional Comments} field allowed them to explain ratings and identify specific strengths and weaknesses. We used the ratings for quantitative comparisons and the comments, where provided, to interpret the observed differences.
Because two evaluators independently rated each repository, we additionally assessed inter-rater agreement using Kendall's $\tau_b$~\cite{kendall1938new,kendall1945ties}. This coefficient measures whether the two evaluators rank the five candidate documents of a repository in a similar order. For each rating dimension, we used the two evaluators' mean rating for each method and repository as the paired observation ($n=20$), compared \textsc{ContribAlign} with each alternative using two-sided Wilcoxon signed-rank tests~\cite{wilcoxon1945individual}, and applied Holm correction~\cite{holm1979simple} to the four comparisons per dimension.

\subsection{Results}
\subsubsection{RQ1: Contribution Guidance Classification}
To answer RQ1, we first examine the validation performance of different model and input configurations and then compare the selected classifier with the LinearSVC classifier selected by Fronchetti et al.~\cite{classify} on the test set.

\begin{wraptable}{r}{0.46\linewidth}
    \centering
    \caption{Weighted F1 Scores of the Five Base Models on the Validation Set}
    \label{tab:ptm-val-grid}
    \footnotesize
    \setlength{\tabcolsep}{2pt}
    \begin{tabular}{@{}lccc@{}}
        \toprule
        Model & \makecell{Only\\paragraph} & \makecell{Paragraph\\+ title} & \makecell{Paragraph + title\\+ context} \\
        \midrule
        BERT-base & 0.717 & 0.748 & 0.758 \\
        ALBERT    & 0.698 & 0.722 & 0.742 \\
        RoBERTa   & 0.730 & 0.752 & \textbf{0.780} \\
        DeBERTa   & 0.726 & 0.773 & 0.773 \\
        CodeBERT  & 0.721 & 0.764 & 0.752 \\
        \bottomrule
    \end{tabular}
\end{wraptable}

Table~\ref{tab:ptm-val-grid} reports the validation weighted F1 scores of the five pretrained language models under the three input configurations. Adding the paragraph title consistently improves performance over using the paragraph alone across all five models. The effect of surrounding context varies across models. It further improves BERT-base, ALBERT, and RoBERTa, leaves DeBERTa unchanged, and slightly reduces the performance of CodeBERT. RoBERTa with the paragraph, title, and surrounding context achieves the highest validation weighted F1 score of 0.780 and is therefore selected as the final classifier.

Table~\ref{tab:baseline-vs-final-test} compares the selected classifier with LinearSVC on the test set. Notably, the weighted F1 score of LinearSVC decreases from 0.651 in the original study~\cite{classify} to 0.608 under our data split. This decrease likely reflects the more stringent repository-level split, which prevents paragraphs from the same repository from appearing in both the training and test sets and thereby reduces potential leakage from shared terminology, phrasing, and document structure.

The selected classifier achieves a weighted F1 score of 0.764, representing an absolute improvement of 0.156 over LinearSVC. Given that the inter-annotator agreement reported for the original corpus was 79\%~\cite{classify}, this result indicates a reasonable level of performance relative to the consistency of the human annotations. Weighted precision increases from 0.625 to 0.767, while weighted recall increases from 0.626 to 0.764. The selected classifier also achieves a higher F1 score for every contribution category and for paragraphs assigned to no category.

\begin{table}[b]
    \centering
    \caption{Test-Set Classification Performance of Fronchetti et al.'s Classifier~\cite{classify} (LinearSVC) and Our Final Classifier (RoBERTa)}
    \label{tab:baseline-vs-final-test}
    \footnotesize
    \begin{tabular}{lrrrrrr}
        \toprule
        & \multicolumn{3}{c}{\makecell[c]{LinearSVC\\(Paragraph)}}
        & \multicolumn{3}{c}{\makecell[c]{RoBERTa\\(Paragraph + Title + Context)}} \\
        \cmidrule(lr){2-4} \cmidrule(lr){5-7}
        Category & F1 & Precision & Recall & F1 & Precision & Recall \\
        \midrule
        Build Local Workspace     & 0.402 & 0.600 & 0.302 & 0.813 & 0.861 & 0.771 \\
        Deal with the Code        & 0.560 & 0.652 & 0.490 & 0.765 & 0.785 & 0.746 \\
        Talk to the Community     & 0.567 & 0.711 & 0.472 & 0.667 & 0.738 & 0.608 \\
        Submit the Changes        & 0.681 & 0.655 & 0.710 & 0.815 & 0.834 & 0.796 \\
        Choose a Task             & 0.272 & 0.824 & 0.163 & 0.500 & 0.648 & 0.407 \\
        Contribution Flow         & 0.364 & 0.492 & 0.288 & 0.582 & 0.557 & 0.608 \\
        \midrule
        No Category Identified  & 0.695 & 0.615 & 0.800 & 0.785 & 0.755 & 0.818 \\
        \midrule
        \textbf{Overall} & \textbf{0.608} & \textbf{0.625} & \textbf{0.626} & \textbf{0.764} & \textbf{0.767} & \textbf{0.764} \\
        \bottomrule
        \vspace{-0.2cm}
    \end{tabular}
\end{table}

The largest improvement is observed for \textit{Build Local Workspace}, whose F1 score increases from 0.402 to 0.813, accompanied by a recall increase from 0.302 to 0.771. Substantial improvements are also observed for \textit{Choose a Task} and \textit{Contribution Flow}, whose F1 scores increase from 0.272 to 0.500 and from 0.364 to 0.582, respectively. Nevertheless, \textit{Choose a Task} remains the lowest-performing category. For this category, the selected classifier improves recall from 0.163 to 0.407 while precision decreases from 0.824 to 0.648, indicating that it identifies a larger proportion of relevant paragraphs at the cost of more false-positive predictions.

\vspace{1mm}
\begin{mdframed}[
    linecolor=gray,
    roundcorner=12pt,
    backgroundcolor=gray!15,
    linewidth=3pt,
    innerleftmargin=2pt,
    leftmargin=0cm,
    rightmargin=0cm,
    topline=false,
    bottomline=false,
    rightline=false,
    skipabove=10pt,
    skipbelow=10pt
]
\textbf{Finding 1:} The selected RoBERTa classifier, using the paragraph, title, and surrounding context, achieves a weighted F1 score of 0.764, an absolute improvement of 0.156 over LinearSVC. It also obtains a higher F1 score in every category and is therefore used for category-aware guidance diagnosis in \textsc{ContribAlign}.
\end{mdframed}
\vspace{1mm}

\subsubsection{RQ2: Evidence Verification}

\begin{wraptable}{r}{0.46\linewidth}
  \centering
  \caption{Confusion Matrix of Verifier and Human Labels}
  \label{tab:rq2-confusion-matrix}
  \footnotesize
  \setlength{\tabcolsep}{3pt}
  \begin{tabular}{lrrr}
    \toprule
    & \multicolumn{3}{c}{Human label} \\
    \cmidrule(l){2-4}
    Verifier label & Supported & Conflicting & Unverified \\
    \midrule
    Supported   & 294 & 0  & 0  \\
    Conflicting & 1   & 36 & 0  \\
    Unverified  & 104 & 2  & 50 \\
    \bottomrule
  \end{tabular}
\end{wraptable}

Table~\ref{tab:rq2-confusion-matrix} compares the verifier's decisions with the human labels on these 487 annotated fragments using a confusion matrix. Counts on the diagonal indicate agreement, while counts off the diagonal indicate deviations by the verifier.
Table~\ref{tab:rq2-class-metrics} reports precision and recall for all three labels. Because sampling was stratified by verifier-assigned label, precision is computed within each prediction stratum and is unchanged by stratum weighting. We report recall both directly from the annotated sample (Sample recall) and after weighting each stratum by its population-to-sample ratio (Estimated recall).

The precision of the \textit{conflicting} classification was 97.3\% (36/37). The sole false positive came from the \texttt{ampproject/amphtml} repository. Its documentation refers to the \texttt{GFI Claimed!} issue label, which exists in the repository's label data. During link extraction, the verifier removed the trailing exclamation mark and instead checked for a nonexistent label named \texttt{GFI Claimed}, leading it to conclude that the documented label did not exist and classify the fragment as \textit{conflicting}.

\begin{wraptable}{r}{0.46\linewidth}
  \centering
  \caption{Precision and Recall by Verification Label}
  \label{tab:rq2-class-metrics}
  \footnotesize
  \setlength{\tabcolsep}{3pt}
  \begin{tabular}{lccc}
    \toprule
    Label & Precision
          & \makecell{Sample\\recall}
          & \makecell{Estimated\\recall} \\
    \midrule
    Supported   & 100.0\% & 73.7\%  & 87.6\% \\
    Conflicting & 97.3\%  & 94.7\%  & 91.5\% \\
    Unverified  & 32.1\%  & 100.0\% & 100.0\% \\
    \bottomrule
  \end{tabular}
\end{wraptable}

The recall of the \textit{conflicting} classification was 94.7\% (36/38) in the annotated sample and 91.5\% after stratum weighting.
Among the fragments classified as \textit{unverified} by the verifier, two were manually confirmed as \textit{conflicting}. This reflects the conservative judgment strategy of the verifier, which does not declare a conflict unless the evidence is definitive.
In the Jetpack guide, \texttt{pnpm lint:fix} does not explicitly include the \texttt{run} subcommand or identify the package in which it should be executed. The verifier therefore could not conclude from the absence of the script alone that the instruction was invalid. Manual inspection found no \texttt{lint:fix} script in either the relevant Boost package or the repository root. In the Etcher guide, the reference to a \texttt{Makefile} does not specify a repository-relative location, so its absence was treated as inconclusive by the verifier. Manual inspection found no \texttt{Makefile} in the examined repository snapshot. No manually confirmed conflicts were found among the sampled fragments classified as \textit{supported}.

The \textit{unverified} stratum contains 104 fragments that the annotator labeled as \textit{supported}. For most of these fragments, the verifier could not establish support through its automated checks, although manual inspection confirmed them as \textit{supported}. For example, one fragment instructs contributors to write a process ID to the \texttt{devices/tasks} file, a virtual cgroup v1 control file rather than an actual file in the repository. Unable to establish its repository-relative root, the verifier marked the fragment as \textit{unverified}, while the annotator confirmed the description as correct. Accordingly, the precision of the \textit{unverified} classification was 32.1\% (50/156), the estimated recall of \textit{supported} was 87.6\%, and the sampled \textit{supported} fragments had a precision of 100.0\%. These disagreements do not exclude the correct fragments from the generation stage, because evidence labeled either \textit{supported} or \textit{unverified} is retained and passed to generation.

\vspace{1mm}
\begin{mdframed}[
    linecolor=gray,
    roundcorner=12pt,
    backgroundcolor=gray!15,
    linewidth=3pt,
    innerleftmargin=2pt,
    leftmargin=0cm,
    rightmargin=0cm,
    topline=false,
    bottomline=false,
    rightline=false,
    skipabove=10pt,
    skipbelow=10pt]
\textbf{Finding 2:} The verifier applies a conservative filtering strategy, marking fragments as \textit{conflicting} only when repository evidence establishes a contradiction and retaining uncertain cases as \textit{unverified}. It achieves 97.3\% precision and 91.5\% estimated recall for conflicts. In the annotated sample, one supported fragment was incorrectly flagged and two actual conflicts remained \textit{unverified}.
\end{mdframed}
\vspace{1mm}

\subsubsection{RQ3: Generated Guideline Quality}

\begin{wraptable}{r}{0.46\linewidth}
    \centering
    \caption{Document Length and Readability}
    \label{tab:rq3-length-readability}
    \footnotesize
    \setlength{\tabcolsep}{2pt}
    \begin{tabular}{@{}lrr@{}}
        \toprule
        Method & \makecell{Word count\\Mean/median}
               & \makecell{Flesch\\Mean/median} \\
        \midrule
        Original              & 534.57/463.50 & 51.05/54.05 \\
        Direct LLM            & 745.13/673.00 & 50.87/53.21 \\
        Adapted LARCH         & 660.59/602.00 & 51.35/52.80 \\
        Generic RAG           & 782.24/683.50 & 53.07/54.05 \\
        \textsc{ContribAlign} & 742.14/726.00 & 51.52/51.12 \\
        \bottomrule
    \end{tabular}

\end{wraptable}

\paragraph{Document Length and Readability.}
Table~\ref{tab:rq3-length-readability} reports the mean and median word counts and Flesch Reading Ease scores~\cite{flesch1948new} for the five candidate documents of each repository, aggregated across 94 repositories. Both measures are calculated on English prose after removing code and Markdown syntax. 
\textsc{ContribAlign} and the three baselines revise and enrich the original contribution guidelines, producing longer documents on average. Such expansion is expected, as prior research has identified substantial gaps in the onboarding guidance provided by many projects~\cite{classify, steinmacher2015barriers}. Among the revised documents, Generic RAG has the highest mean word count, while Adapted LARCH has the lowest. Flesch Reading Ease scores are similar across all five document types, suggesting no marked difference in surface readability despite the increase in length.

\paragraph{Human Evaluation Scores and Preferences.}


Table~\ref{tab:human_evaluation} shows that \textsc{ContribAlign} achieves the highest mean score on all dimensions and is selected as the best document in 33 of 40 evaluation cases. It achieves significantly better performance than all four alternatives on usefulness, beginner-friendliness and completeness, and outperforms the three generated baselines in correctness. 
The median Kendall's $\tau_b$ ranged from 0.71 to 0.80 across the four dimensions, indicating generally similar rankings between evaluators and supporting the reliability of the evaluation.

On Correctness, \textsc{ContribAlign} significantly outperforms the three generative baselines. The original document, often written by project maintainers, tends to be accurate and does not differ significantly from \textsc{ContribAlign}. The following examples illustrate the benefits and limits of evidence verification. In \texttt{aframevr/aframe}, the original document and the three generative baselines all advise newcomers to look at the \texttt{help wanted (easy)} and \texttt{help wanted (hard)} labels, but the latter label does not exist in the repository. The verifier classifies this instruction as \textit{conflicting} and filters it out. \textsc{ContribAlign} then generates guidance that recommends only \texttt{help wanted (easy)}, based on the repository's labels and open issue records. Nevertheless, the verifier does not eliminate all errors. In \texttt{aaubry/YamlDotNet}, the \textsc{ContribAlign} document still retains an outdated list of target frameworks that was classified as \textit{unverified}. These two cases show that identified conflicts can be filtered out, whereas retained unverified content still requires careful scrutiny.

\begin{table}[t]
\centering
\caption{
Human evaluation results across 20 repositories, each assessed by two evaluators.
Scores are mean Likert ratings (1--5) over 40 evaluations per method.
Preference is the number (\%) of evaluator--repository pairs selecting each
document as the best.
\textit{Note.} Asterisks indicate differences from \textsc{ContribAlign}
based on two-sided paired Wilcoxon signed-rank tests using 20 pairs of repository-level mean ratings, with Holm correction applied separately to the four comparisons in each dimension:
$^{*}p<0.05$, $^{**}p<0.01$, and $^{***}p<0.001$.
}
\label{tab:human_evaluation}
\footnotesize
\renewcommand{\arraystretch}{1.08}
\begin{tabular*}{0.92\linewidth}{@{\extracolsep{\fill}}lccccc@{}}
\toprule
Method
& Correctness
& Usefulness
& \makecell{Beginner-\\friendliness}
& Completeness
& Preference \\
\midrule
\textbf{\textsc{ContribAlign}}
& \textbf{4.475}
& \textbf{4.550}
& \textbf{4.325}
& \textbf{4.325}
& \textbf{33/40 (82.5\%)} \\
Generic RAG
& \sig{3.825}{**}
& \sig{3.750}{**}
& \sig{3.600}{**}
& \sig{3.775}{*}
& 3/40 (7.5\%) \\
Direct LLM
& \sig{3.600}{***}
& \sig{3.550}{***}
& \sig{3.600}{**}
& \sig{3.525}{**}
& 0/40 (0\%) \\
Adapted LARCH
& \sig{3.675}{***}
& \sig{3.050}{***}
& \sig{3.325}{**}
& \sig{3.075}{***}
& 3/40 (7.5\%) \\
Original
& 4.000
& \sig{2.650}{***}
& \sig{2.850}{**}
& \sig{2.525}{***}
& 1/40 (2.5\%) \\
\bottomrule
\end{tabular*}
\end{table}

\textsc{ContribAlign} significantly outperforms the other four document versions in both Usefulness and Beginner-friendliness. Evaluator comments indicate that actionable guidance for environment setup, validation, and contribution submission is particularly important. For example, the original document of \texttt{android-async-http/android-async-http} was described as \textit{``pure policy text without any local environment setup, build commands, or version requirements.''} In contrast, \textsc{ContribAlign} provides repository-consistent Gradle, SDK, and CI configurations with accurate validation commands, enabling newcomers to \textit{``set up the environment in three steps.''}

On Completeness, \textsc{ContribAlign} also significantly outperforms the other four document versions, while the original documents receive the lowest mean score. Evaluators praised its \textit{``complete coverage of community entry points, task labels, workspace, CI, and the submission process''} in \texttt{alibaba/formily}, and its coverage of the current environment, installation, community, and pull request workflows in \texttt{cvxgrp/cvxpy}. These comments suggest that completeness depends on providing useful guidance across multiple stages of contribution, rather than simply producing longer documents. Generic RAG also retrieves repository documents for all six categories and thus provides broader coverage than the other generative baselines. However, retrieval alone may not provide the category-specific, actionable information needed at each stage, which may help explain why Generic RAG still significantly underperforms \textsc{ContribAlign}.

Overall preferences are consistent with these ratings, with \textsc{ContribAlign} preferred in 33 of 40 evaluation records and by both evaluators in 15 of 20 repositories. In \texttt{actions/cache}, neither evaluator preferred \textsc{ContribAlign}; both selected Generic RAG for its more balanced coverage and noted that \textsc{ContribAlign} omitted the security-reporting channel. Other exceptions reflect evaluator-specific trade-offs. For \texttt{cnvs/canvas}, one evaluator preferred Adapted LARCH for its concise and balanced guidance, while the other preferred the original for its clear workflow and directly executable commands, finding \textsc{ContribAlign} more complete but overly long. These cases suggest that overall preference reflects trade-offs among completeness, correctness, conciseness, and organization, and that \textsc{ContribAlign} can still contain omissions or less effective guidance despite its overall advantage.

\vspace{1mm}
\begin{mdframed}[
    linecolor=gray,
    roundcorner=12pt,
    backgroundcolor=gray!15,
    linewidth=3pt,
    innerleftmargin=2pt,
    leftmargin=0cm,
    rightmargin=0cm,
    topline=false,
    bottomline=false,
    rightline=false,
    skipabove=10pt,
    skipbelow=10pt]
\textbf{Finding 3:} \textsc{ContribAlign} performs best overall, achieving the highest mean scores on all four dimensions and being selected as the best document in 33 of 40 evaluator--repository assessments. It significantly improves usefulness, beginner-friendliness, and completeness, while maintaining correctness comparable to the original documents and higher than the three generative baselines.
\end{mdframed}
\vspace{1mm}

\section{Discussion}

\subsection{Understanding Contribution Guidance Beyond Broad Categories}
The fine-grained taxonomy developed in this work is not only a component of \textsc{ContribAlign}, but also an empirical characterization of what \texttt{CONTRIBUTING} files contain and what they may need to contain. Prior work organizes contribution guidance into six contribution categories, which is useful for identifying whether a document addresses major stages of the contribution process. However, these broad categories do not explain what concrete information should appear within each category. For example, knowing that a document should include \textit{Talk to the Community} guidance does not indicate whether it should describe available communication channels, the purposes of different channels, how contributors can ask for help, expected interaction norms, or the Code of Conduct.

By analyzing annotated \texttt{CONTRIBUTING} content, our taxonomy makes these internal information needs explicit. This provides a more detailed view of contribution guidance than category-level coverage alone. It can therefore serve as a checklist for maintainers who manually write or revise contribution guidelines: rather than only asking whether a document contains a section about task selection, local setup, or submission, maintainers can inspect whether the document covers the fine-grained types that commonly support those tasks. In \textsc{ContribAlign}, the taxonomy also provides structured information needs for evidence collection and category-wise generation. More broadly, it contributes a reusable description of the content structure of \texttt{CONTRIBUTING} files, independent of any particular generation pipeline.

\subsection{Balancing Conciseness and Detail for Newcomers}
A recurring tension in revising contribution guidance is the balance between conciseness and detail. Newcomers do not necessarily have uniform preferences. 
Some may prefer a concise document that quickly points them to the main contribution workflow, communication channels, and submission rules. Others may need more detailed instructions, especially when they are unfamiliar with the project's technology stack, build process, testing expectations, or review conventions. A short document may be easier to scan, but it can leave important operational questions unanswered. A detailed document can reduce uncertainty, but it may also increase the amount of information that newcomers must read before making progress.
This tension suggests that there may not be a single universally optimal form of a \texttt{CONTRIBUTING} file. \textsc{ContribAlign} currently favors evidence-supported and relatively complete guidance, because the goal is to repair missing or outdated information in existing files. This strategy is useful when the original document lacks concrete instructions or when project-specific procedures are scattered across repository materials. However, in some settings, a more concise version may be preferable, especially for experienced contributors or projects with simple workflows. Future systems could support different revision strategies, such as a concise mode that emphasizes essential steps and links to detailed materials, and a detailed mode that provides fuller evidence-supported instructions for newcomers who need more guidance. Such customization would better reflect the diverse information needs of OSS contributors.

\section{Threats to Validity}


\subsection{Construct Validity}

We assess document quality along four dimensions: correctness, usefulness, beginner-friendliness, and completeness. These dimensions are not strictly independent. 
We therefore interpret the scores as complementary perspectives on document quality rather than as measurements of orthogonal constructs. In addition, the ratings reflect evaluators' judgments rather than newcomers' direct use of the documents. They may therefore not fully capture newcomers' experiences during actual contribution tasks. To partially mitigate this threat, we recruited computer science graduate students with no prior contributions to the evaluated projects to provide a project-newcomer perspective.

\subsection{Internal Validity}

Category-aware diagnosis relies on an automatic classifier, and classification errors may affect category coverage judgments and the organization of original guidance for revision. Another threat arises from the scope of fact verification. The verifier retains inconclusive fragments as \textit{unverified}, while some contribution practices fall outside its checking scope. Errors in retained content may therefore propagate into revised documents. Given this risk, we position the revised document as a proposed alternative to the existing \texttt{CONTRIBUTING} file, which community maintainers should review before adoption.

We evaluate the classifier and verifier independently and compare the complete pipeline with the baselines at the document level. Without end-to-end component ablations, these evaluations cannot isolate each component's contribution to document quality. We therefore interpret the quality of the revised documents as the outcome of the complete pipeline rather than the contribution of any individual component.




\subsection{External Validity}
Our document generation evaluation uses 94 popular GitHub repositories from the classifier's test set in the manually annotated corpus of Fronchetti et al.~\cite{classify}. We excluded inaccessible repositories, repositories without a retrievable current \texttt{CONTRIBUTING} file, and archived repositories. Our results therefore mainly reflect relatively mature GitHub projects.

The effectiveness of \textsc{ContribAlign} may vary with project characteristics such as size, activity, and hosting platform. Projects that differ from the studied repositories in these respects usually provide less evidence for retrieval, verification, and generation, and the coverage of configuration-based and execution-based evidence also differs across programming languages and software ecosystems. Automatic revision may therefore be less effective and may require more direct input from maintainers. 

The document-level human evaluation covers only 20 repositories, approximately one fifth of the 94 target repositories. Although the repositories were stratified by primary programming language to cover different project ecosystems, the representativeness of the evaluation results is limited by this sample size.

In addition, all generation methods rely on the same LLM (DeepSeek-V4.1-Flash). Because the model is held constant across methods, comparisons between methods are not confounded by model choice. However, our conclusions are tied to the capabilities of this particular model. With a different model, the magnitude of \textsc{ContribAlign}'s advantage over the baselines could change. Future work should evaluate \textsc{ContribAlign} on more diverse projects and generation models.

\section{Conclusion}
This paper presents \textsc{ContribAlign}, a framework for automatically revising \texttt{CONTRIBUTING} files through category-guided evidence integration. \textsc{ContribAlign} diagnoses category coverage in existing guidance, collects evidence from repository documents, configurations, and collaboration activities, and verifies repository-specific claims before generating revisions category by category. Evaluation demonstrates that the classifier improves category diagnosis over prior work, the verifier detects conflicting fragments with high precision, and the revised documents receive higher human ratings than the original files and three baselines. Our findings highlight the potential of grounding document revision in category-specific repository evidence while filtering detected conflicts.

\section{Data Availability}
The ContribAlign replication package, including the implementation, evaluation materials, and results, is available at \url{https://figshare.com/s/d23c57c418fc0668f0a5}. Upon acceptance, we intend to make the replication package publicly available.

\bibliographystyle{ACM-Reference-Format}
\bibliography{references}

\appendix

\end{document}